\documentclass[prd,twocolumn,showpacs,nofootinbib, aps,10pt,superscriptaddress,amsmath,amssymb]{revtex4-1}

\usepackage{graphicx}% Include figure files
\usepackage{dcolumn}% Align table columns on decimal point
\usepackage{bm}% bold math
\usepackage{epstopdf}
\usepackage{mathtools}
\usepackage{natbib}
\usepackage{captcont}
\usepackage{booktabs}
\usepackage{threeparttable}
\usepackage{slashed}
\usepackage{color}

\begin{document}
\title{Evidence of a simple dark sector from XENON1T excess}

\author{Cheng-Wei Chiang}
\email{chengwei@phys.ntu.edu.tw}
\affiliation{Department of Physics, National Taiwan University, Taipei, Taiwan 10617, Republic of China}

\author{Bo-Qiang Lu}
\email{bqlu@phys.ntu.edu.tw}
\affiliation{Department of Physics, National Taiwan University, Taipei, Taiwan 10617, Republic of China}

\begin{abstract}

We propose that the nearly massless dark photons produced from the annihilation of keV dark fermions in the Galaxy can induce the excess of 
electron recoil events recently observed in the XENON1T experiment.  The minimal model for this is the extension of a $U(1)_{X}$ gauge symmetry, 
under which the dark photon couples to both dark and visible matter currents.  We find that the best-fit parameters of the dark sector are 
compatible with the most stringent constraints from stellar cooling.  
We also show that in the freeze-out scenario, the dark fermions can explain the anomaly while contributing $\agt 1\%$ of the DM relic density.

\end{abstract}
\maketitle

% {\it Introduction ---}
\section{Introduction}
One of the outstanding puzzles in particle physics is the nature of dark matter (DM), whose existence has been confirmed from various cosmological 
and astrophysical observations~\cite{Planck2016AA}.  An attractive hypothesis, the weakly interacting massive particles (WIMPs)~\cite{Lee1977PRL} 
having mass at around the electroweak scale and coupling strength similar to the weak coupling, have suffered stringent constraints from both 
direct~\cite{XENON1T2018} and indirect~\cite{Ackermann2015PRL} DM detection experiments.  As alternatives to the WIMP DM, sub-GeV or (super-)light 
DM candidates, such as axion and dark photon, have drawn more attention in recent years.

The result of searches for new physics using low-energy electronic recoil data recently published by XENON Collaboration shows an excess of 
events over the known backgrounds in the recoil energy range 1--7~keV, peaked around 2.4~keV, with a local statistical significance of 
3--$4\sigma$~\cite{XENON1T2020}.  A series of theoretical works have been inspired to explain the origin of the excess 
(see Ref.~\cite{Bloch2020} for a summary).
As noted by the XENON Collaboration, the absorption of axions emitted by the Sun can fit the data quite well.  
However, the favored parameter space is in severe tension with stellar cooling constraints~\cite{XENON1T2020}.  
If axion-like particles (ALPs) are assumed to be the sole DM that couple predominantly to electrons (the so-called photophobic models), 
the XENON1T anomaly can be explained without being excluded by the constraints~\cite{Bloch2020}.  
The Dine-Fischler-Srednicki-Zhitnitsky~\cite{DFSZ} type of ALPs are excluded by the stellar cooling since they strongly couple to both 
electrons and photons. The XENON1T experiment is in general insensitive to the Kim-Shifman-Vainshtein-Zakharov~\cite{KSVZ} type of ALPs, which 
couple to an electron at the loop level~\cite{XENON1T2020}. The absorption of massive dark photon DM also suffer strong constraints from 
stellar cooling observations~\cite{Alvarez2020}.  
Boosted DM (BDM) particles that have velocities $\sim 0.06c$ can lead to a keV-scale energy deposition via BDM-electron scattering~\cite{AcceleratedDM}.
Room for the galactic BDM explanation is still available when various limits are taken into account~\cite{Alhazmi2020}, while the Sun cannot be the source of the BDM 
flux because of its large cross section to be trapped inside the Sun~\cite{Fornal2020}.

In this work, we propose the dark fermion annihilation to nearly massless dark photons in the Galaxy as the 
origin for the XENON1T anomaly.  
As shown below, our scenario is quite simple and natural, and can survive the most stringent constraints from stellar cooling observations.
Furthermore, the XENON1T data can be accommodated even when the dark fermions constitute only a small fraction of the DM.

% {\it Dark Sector ---}
\section{Dark Sector}
Consider a simple extension of the Standard Model (SM) with a dark $U(1)_X$ gauge symmetry, with the gauge field denoted by $\tilde{X}_{\mu}$, 
and a Dirac dark fermion $\chi$ carrying $U(1)_X$ charge $q_X = 1$.  SM particles are all neutral under the $U(1)_X$ gauge symmetry.  
The relevant Lagrangian describing the dark sector, the photon field, and the mixing between the visible and dark photons reads
\begin{equation}
    \label{eq:u1larg}
    \mathcal{L}_0=-\frac{1}{4}X_{\mu \nu}X^{\mu \nu}-\frac{1}{4}F_{\mu \nu}F^{\mu \nu}-\frac{\varepsilon}{2}X_{\mu \nu}F^{\mu \nu}
    +\bar{\chi}(i\slashed{D} - m_{\chi})\chi
    ~,
\end{equation}
where $X_{\mu \nu}=\partial_{\mu} \tilde{X}_{\nu}-\partial_{\nu} \tilde{X}_{\mu}$ and $F_{\mu \nu}=\partial_{\mu} \tilde{A}_{\nu}-\partial_{\nu} \tilde{A}_{\mu}$, 
and the covariant derivative $D_{\mu}=\partial_{\mu}-i e_{X} q_{X} \tilde{X}_{\mu}$, with $e_X$ denoting the dark gauge coupling.  
The parameter $\varepsilon\ll 1$ represents the kinetic mixing between the dark gauge boson $\tilde{X}_\mu$ and the $U(1)_{\rm em}$ gauge 
boson $\tilde{A}_{\mu}$.

The kinetic terms of gauge boson in the Eq.~\eqref{eq:u1larg} can be diagonalized by rotating the gauge fields as
\begin{equation}
    \left(\begin{array}{c}
    \tilde{X}_{\mu} \\
    \tilde{A}_{\mu}
    \end{array}\right)=\left(\begin{array}{cc}
    1 & 0 \\
    -\varepsilon & 1
    \end{array}\right)\left(\begin{array}{cc}
    \cos \theta & -\sin \theta \\
    \sin \theta & \cos \theta
    \end{array}\right)\left(\begin{array}{c}
    X_{\mu} \\
    A_{\mu}
\end{array}\right)
~,
\end{equation}
where we now identify $A_{\mu}$ as the visible photon and $X_{\mu}$ as the dark photon with mass $m_{\gamma'}$.
The rotation angle $\theta$ would be locked at zero if the dark photon becomes massive from gauge symmetry breaking~\cite{Fabbrichesi2020}.
In this case, the nearly massless (compared to the keV scale) dark photon $X_{\mu}$ couples to both visible and dark currents while the 
ordinary photon $A_{\mu}$ couples exclusively to the visible current:
\begin{equation}
    \label{eq:largi}
    \mathcal{L}=eJ^{\mu}A_{\mu}+\left (e_{X}{J}_{X}^{\mu}-\varepsilon eJ^{\mu} \right )X_{\mu}
    ~,
\end{equation}
where $e$ is the electric charge, $J^{\mu}$ is the visible current, and ${J}_{X}^{\mu}=\bar{\chi}\gamma^{\mu}\chi$ is the dark current.

Our interpretation for the XENON1T excess is the following.  The dark fermions $\chi$ pair annihilate into the dark photon pairs via the 
$t$ and $u$-channel exchanges, and the dark photons are absorbed by the detector material and lead to the electron emissions because of the dark 
photoelectric effect.  Since the dark fermions are non-relativistic, each of the dark photons thus produced has an energy $E_{{\gamma}'}=m_{\chi}$.  
Furthermore, the dark fermion annihilation into other SM particles is kinematical forbidden when $m_{\chi}<m_e=511$~keV.  Note that the mean free 
path of keV visible photons in the atmosphere is $l_{\gamma}=1/(\rho_{\rm gas}\sigma_{\gamma}^{\rm gas}(\rm keV))\sim 1~{\rm cm}$ due to strong 
absorption.  Hence, such photons produced in the space would not be able to reach the detector on Earth.  In contrast, the mean free path of keV 
dark photon can be estimated to be $l_{{\gamma}'}=l_{\gamma}/\varepsilon^2$.
Provided that the mixing parameter $\varepsilon\sim 10^{-10}$, the mean free paths for keV dark photon propagating through the atmosphere and Earth 
are, respectively, $\sim 10^{20}$~cm and $10^{16}$~cm, much larger than the Earth radius. 
Thus, the atmosphere and Earth are transparent to the keV dark photons.

% {\it XENON1T Excess ---}
\section{XENON1T Excess}
The differential dark photon flux from dark fermion annihilation is given by~\cite{Ackermann2015PRL}
\begin{equation}
    \label{eq:DMdflux}
    \frac{d \Phi_{{\gamma}'}}{d E_{{\gamma}'}}=\frac{\langle\sigma v\rangle_{{\gamma}'{\gamma}'}}{2m_{\chi}^{2}} \frac{d N_{{\gamma}'}}{d E_{{\gamma}'}}J_{\chi}
    ~,
\end{equation}
where $\langle\sigma v\rangle_{{\gamma}'{\gamma}'}$ denotes the thermally averaged annihilation cross section of dark fermions to a pair of dark photons
and the dark fermion {\it J}-factor is $J_{\chi}=\frac{1}{4\pi}\int d\Omega \int ds \rho_{\chi}^2$.  Suppose the dark fermions constitute a 
fraction of the observed DM, {\it i.e.}, the dark fermion density and the DM mass density are related by $\rho_{\chi}=f_{\chi}\rho_{\rm DM}$ 
with $f_\chi \le 1$.  Here we assume the NFW profile~\cite{NFW1997} for the Galactic DM halo profile,
\begin{equation}
    \rho_{\rm DM}(r)=\frac{\rho_s}{(r/r_s)(1+r/r_s)^2}
    ~,
\end{equation}
where $r_s=20$~kpc and $\rho_{s}=0.26~{\rm GeV~cm^{-3}}$, corresponding to a local DM mass density $0.3~{\rm GeV~cm^{-3}}$. 
The dark fermion {\it J}-factor is then determined by 
$J_{\chi}=f_{\chi}^2J_{\rm DM}\simeq f_{\chi}^2\times 10^{22}~{\rm GeV^{2}~cm^{-5}}$~\cite{Ackermann2015PRL}.  
The dark photon energy spectrum is
\begin{equation}
    \frac{d N_{{\gamma}'}}{d E_{{\gamma}'}}=2\delta(E_{{\gamma}'}-m_{\chi})
    ~.
\end{equation}
Since the annihilation can proceed through {\it s}-wave processes, the thermally averaged cross section 
$\langle\sigma v\rangle_{{\gamma}'{\gamma}'}$ at the leading order is given by
\begin{equation}
    \langle\sigma v\rangle_{{\gamma}'{\gamma}'} \simeq \frac{\pi \alpha_X^{2}}{2m_{\chi}^2}+\mathcal{O}(v^2)
    ~,
\end{equation}
with $\alpha_X \equiv e_X^2/(4\pi)$.

%--------------------------------------------------------
\begin{figure}[t]
\vspace{-0.5cm}
    \includegraphics[width=95mm,angle=0]{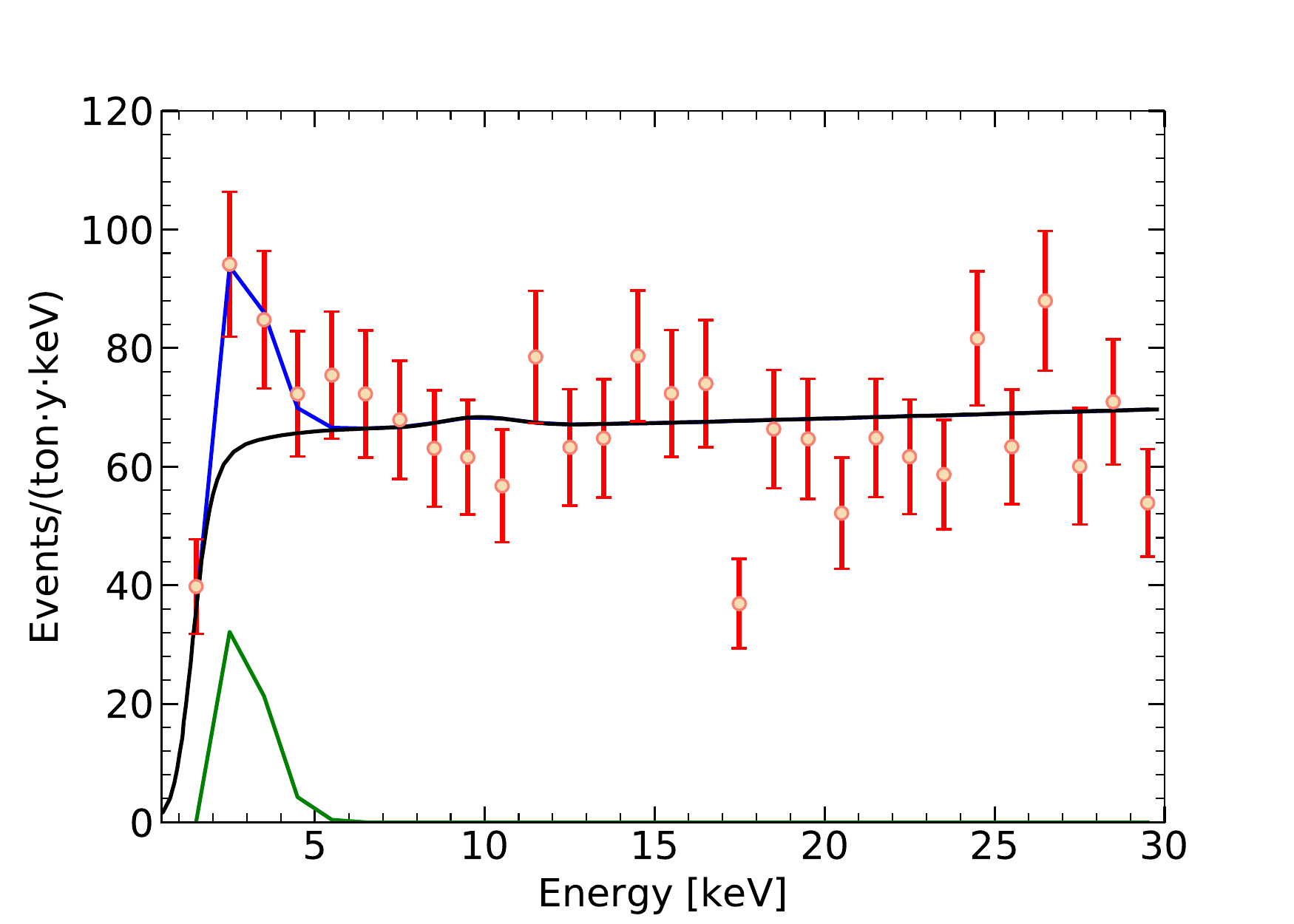}
    \caption{The energy spectrum of recoiled electrons for the benchmark of $m_{\chi}=3.17$~keV and ${\alpha}'=1.46\times 10^{-21}$, represented by the blue curve.  The data and background (black curve) are extracted from Ref.~\cite{XENON1T2020}.  The green curve represents the contribution from the absorption of dark photons.}
    \label{fig:ann}
\vspace{-0.3cm}
\end{figure}
%---------------------------------------------------------

The event rate due to the absorption of dark photons produced from dark fermion annihilation in the detector is given by~\cite{Armengaud2013JCAP}
\begin{eqnarray}
    \label{eq:absortR1}
    \frac{dR(E)}{d\Delta E}&=&\int dE_{{\gamma}'}\sigma_{{\gamma}'}(E_{{\gamma}'})\frac{d\Phi_{{\gamma}'}}{dE_{{\gamma}'}}\epsilon (E)MT
    \frac{1}{\sqrt{2\pi}\sigma}e^{-\frac{(E-E_{{\gamma}'})^2}{2\sigma^2}}\nonumber\\
    &=&\frac{\sigma_{{\gamma}'}(m_{\chi})}{\sqrt{2\pi}\sigma}\frac{\langle\sigma v\rangle_{{\gamma}'{\gamma}'} }{m_{\chi}^2}J_{\chi}\epsilon (E)MT
    e^{-\frac{(E-m_{\chi})^2}{2\sigma^2}}
    ~,
\end{eqnarray}
where $\sigma_{{\gamma}'}=\varepsilon^2 \sigma_{\gamma}$ is the dark photoelectric cross section of the detector material for photons with 
energy $E_{{\gamma}'}$, with the photoelectric cross section $\sigma_{\gamma}$ given in Ref.~\cite{PhotoelectricCS} 
(the in-medium effects of dark photon absorption can be neglected for $E_{\gamma'}\gg 12.13$~eV~\cite{An2015PLB});
$\epsilon (E)$ is the total efficiency for the XENON1T experiment; $MT= (1042~{\rm kg}) \times (226.9~{\rm days})$ is the exposure; 
and $\sigma$ is the experimental energy resolution~\cite{XENON2019Nature}.  
A fit to the experimental energy resolution data distribution gives $\sigma/E=0.341/\sqrt{E}$.  
With Eq.~\eqref{eq:DMdflux} to Eq.~\eqref{eq:absortR1}, the event rate is then estimated to be
\begin{eqnarray}
\frac{dR}{d\Delta E}=&&1.737\times 10^{40}(f_{\chi}{\alpha}')^2\epsilon (E)\left ( \frac{\rm keV}{m_{\chi}} \right )^4
\left ( \frac{\sigma_{\gamma}(m_{\chi})}{\rm barns} \right )\nonumber\\
&&\times \frac{1}{\sqrt{2\pi}\sigma}e^{-\frac{(E-m_{\chi})^2}{2\sigma^2}}
~,
\end{eqnarray}
where ${\alpha}' \equiv \varepsilon \alpha_X = \varepsilon e_X^2/(4\pi)$.
If we assume that the DM is entirely comprised of the dark fermions, {\it i.e.}, $f_{\chi}=1$, ${\alpha}'\sim 10^{-21}$ is sufficient to 
generate $\sim 10$ photoelectric events in the XENON1T detector, using the xenon photoelectric cross section 
$\sigma_{\gamma}(3~\rm keV)\sim 10^{5}$~barns/atom.  Figure.~\ref{fig:ann} shows the predicted energy spectrum.  
The signal events represented by the green curve are calculated using the benchmark parameters $m_{\chi}=3.17$~keV and 
${\alpha}'=1.46\times 10^{-21}$.  The data points and background events (black curve) are taken from Ref.~\cite{XENON1T2020}.  
The blue curve represents the total events in our model.

%--------------------------------------------------------
\begin{figure}[t]
\vspace{-0.5cm}
    \includegraphics[width=95mm,angle=0]{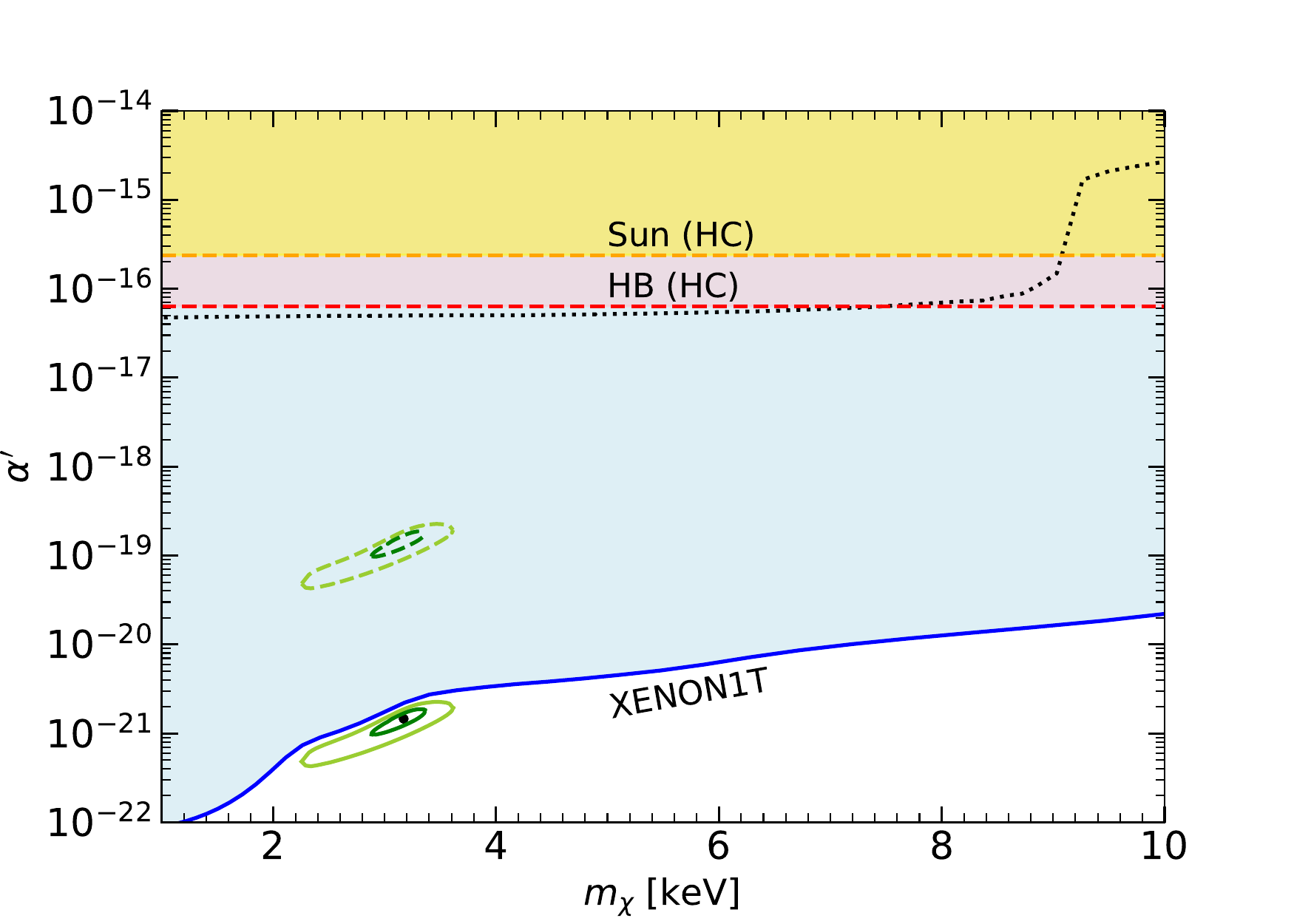}
    \caption{The solid (dashed) green contours represent the parameter region favored by the XENON1T data for $f_{\chi}=1$ ($10^{-2}$) at 
    the $1\sigma$ (darker) and $2\sigma$ (lighter) levels. For $f_{\chi}=1$, the 95\% C.L. upper limits on ${\alpha}'$ as derived from the 
    XENON1T data is represented by the blue curve.  
    The constraints on the HC dark photon from the Sun and HB stars are taken from Ref.~\cite{An2013PRL}, 
    while the black dotted curve represents the RG constraint on the dark fermion from Ref.~\cite{Vogel2014JCAP}. 
    The colored regions are excluded by the constraints.
    }
    \label{fig:cont}
\vspace{-0.3cm}
\end{figure}
%---------------------------------------------------------

In Fig.~\ref{fig:cont}, we depict the parameter region favored by the XENON1T data at $1\sigma$ (dark green curves) and $2\sigma$ 
(light green curves) confidence level (C.L.).  The solid curves are drawn for the scenario when all the DM are explained by the dark fermion, 
whereas the dashed curves are for the scenario when the dark fermion constitutes $1\%$ of the DM relic density.  
To do this, we calculate the signal and background from Ref.~\cite{XENON1T2020} in 29 equidistant bins between 1~keV and 30~keV.  
The goodness of the fit to data is estimated using a $\chi^2$ test.  We find that the best-fit parameters are $m_{\chi}=3.17\pm 0.21~{\rm keV}$ 
and ${\alpha}'=(1.46\pm 0.44)\times 10^{-21}$, marked by the black dot in the figure, with $\chi_{\rm min}^2/{\rm d.o.f}=35.90/27$ 
(corresponding to a $p$-value of $11.75\%$).  A purely background fit to the data gives $\chi_{\rm bkg}^2=46.35/29$ ($p$-value $=2.17\%$).  
We also derive the 95\% C.L. upper limits on the coupling ${\alpha}'$ for a given value of $m_{\chi}$ by increasing the minimum 
$\chi^2$ by $\Delta\chi^2=5.99$.  Such a limit is given by the blue curve, and the light-blue region is excluded at 95\% C.L.

%--------------------------------------------------------
\begin{figure}[t]
\vspace{-0.5cm}
    \includegraphics[width=95mm,angle=0]{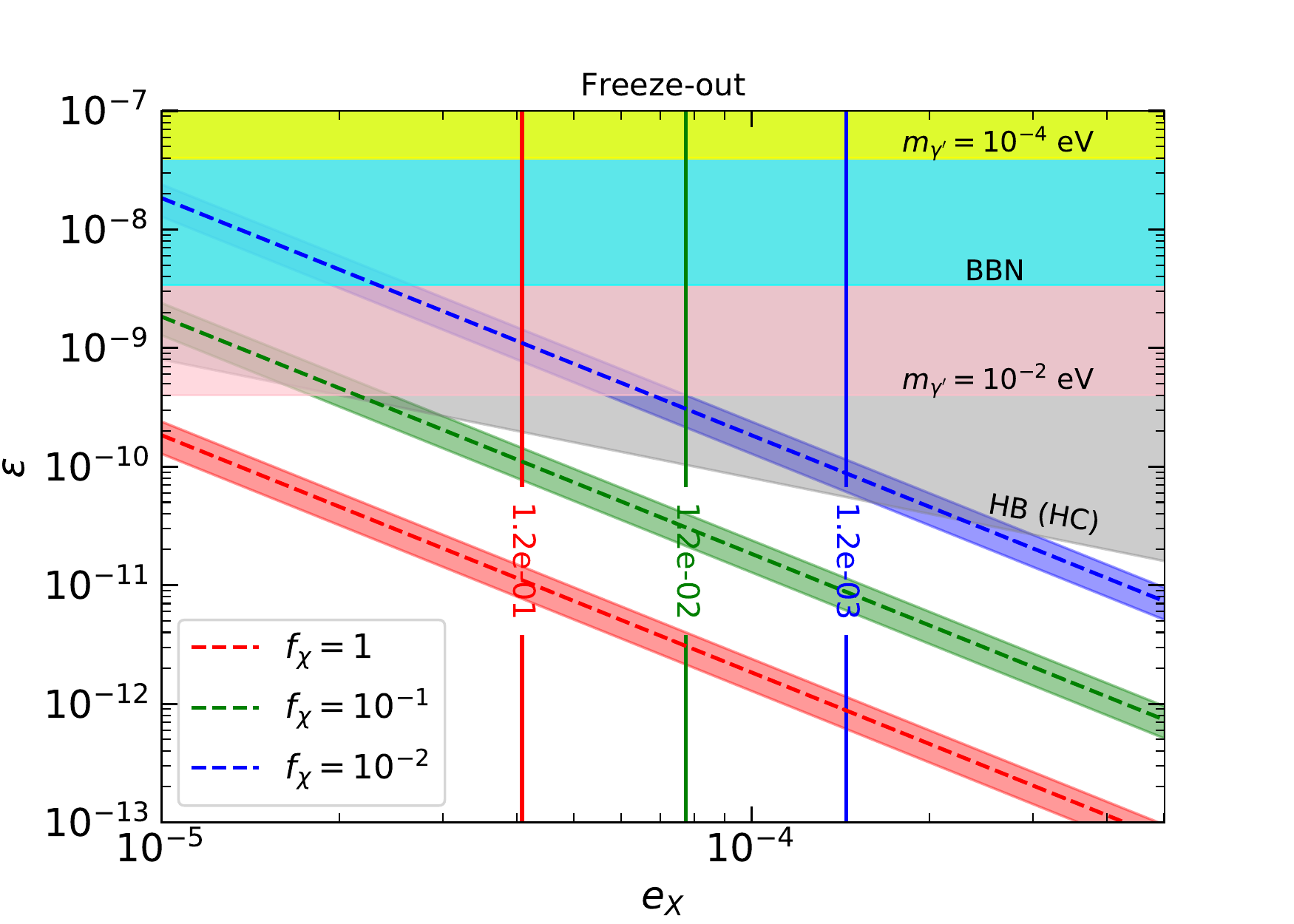}
    (a)
    \includegraphics[width=95mm,angle=0]{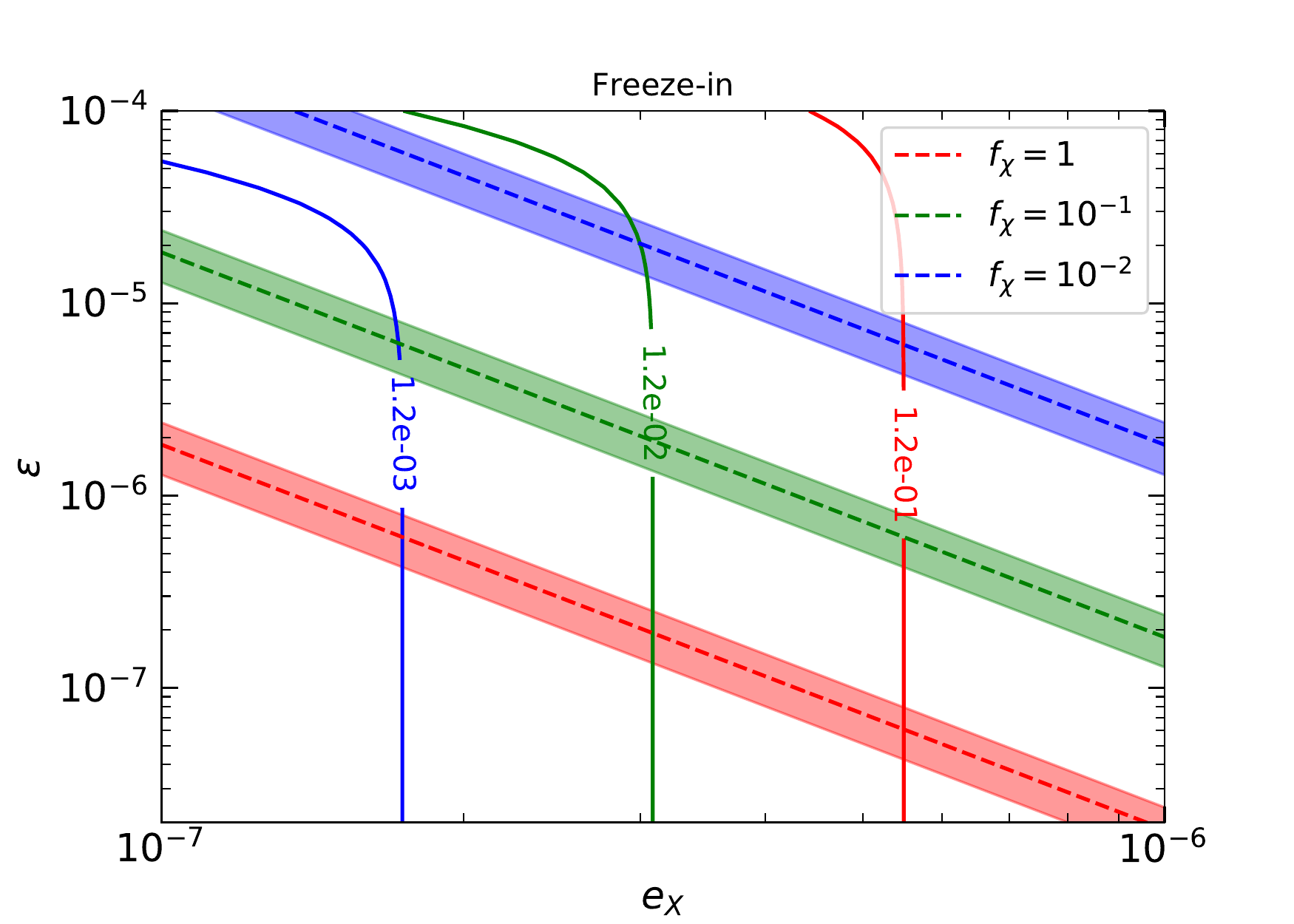}
    (b)
    \caption{The solid curves correspond to the dark fermion relic density $\Omega_\chi h^2 = 0.12$ (red), 0.012 (green), and 0.0012 (blue), 
    and the dashed curves correspond to the best-fit value of ${\alpha}'$ in the corresponding case in the $e_{X}-\varepsilon$ plane. 
    The colored bands represent the 1$\sigma$ uncertainty in ${\alpha}'$.  Plot (a) and plot (b) are for the freeze-out and freeze-in scenarios, 
    respectively. The dark fermion and dark photon masses are fixed, respectively, at $3.17$~keV and $m_{\gamma'}=10^{-20}$~eV in both plots,
    and the relic density is calculated with the help of $\textsf{MicrOMEGAs}\_\textsf{5.0.4}$~\cite{Barducci2018CPC}.
    Plot (a) also shows the limits from BBN, constraints on the HC dark photon from HB stars, and constraints on the SC dark photon from 
    the Sun with $m_{\gamma'}=10^{-2}$~eV and $10^{-4}$~eV, respectively~\cite{An2013PRL}.
    }
    \label{fig:relic}
\end{figure}
%---------------------------------------------------------

% {\it Constraints ---}
\section{Constraints}
The most relevant and stringent constraints for mass scale in the range $\lesssim 10$~keV come from astrophysical observations of stellar cooling processes. 
The black dotted curve in Fig.~\ref{fig:cont} represents the red giants (RG) constraints on dark fermion from Ref.~\cite{Vogel2014JCAP}.  
Note that Ref.~\cite{Vogel2014JCAP} assumes a massless dark photon, and therefore, the SM photon couples to the dark current, which, however, 
does not occur in the current model. We thus take into account the stellar cooling constraints on the massive dark photon as well~\cite{An2013PRL}.  
However, these constraints depend on the origin of dark photon mass, including the Stueckelberg case (SC) and the Higgs case (HC)~\cite{An2013PRL}.  
For the dark photon with mass $m_{\gamma'}\lesssim$~keV, the horizontal branch (HB) stars constraint on the HC dark photon is $\varepsilon e_X<8\times 10^{-15}$ 
(red dashed line in Fig.~\ref{fig:cont}) while the Sun constraint on the SC dark photon is $\varepsilon m_{\gamma'}/{\rm eV}<4\times 10^{-12}$ 
(depicted by the pink and yellow regions in Fig.~\ref{fig:relic} for $m_{\gamma'}=10^{-2}$~eV and $10^{-4}$~eV, respectively)~\cite{An2013PRL}.  
Bounds from cosmological~\cite{McDermott2020PRD} and atomic~\cite{Jaeckel2010PRD} experiments have excluded $\varepsilon\gtrsim 3\times 10^{-8}$ 
for $10^{-14}{~\rm eV}\lesssim m_{\gamma'}\lesssim 10^{-4}{~\rm eV}$.  If the dark sector is in thermal equilibrium with the SM sector, 
it could affect big bang nucleosynthesis (BBN) via altering the effective number of thermally excited neutrino degrees of freedom $N_{\rm eff}$. 
Following Ref.~\cite{Davidson1991PRD}, we obtain the constraint $\varepsilon<3.4\times 10^{-9}$ by requiring that the dark sector decouples from 
the SM sector at $T=2$~MeV, before BBN starts.  We note in passing that owing to the absence of photon pair annihilation in our model at tree level, 
the constraints arising from the observational X-ray data~\cite{Signe2016AA} do not apply.

From Fig.~\ref{fig:cont} we observe that the regions (with $f_{\chi}=1$) inferred from the XENON1T experiment are far below the constraints from stellar cooling.  
In fact, the allowed parameter space is so large that even if the dark fermion constitutes only a small fraction of DM, 
the dark fermion annihilation can still possibly account for the XENON1T excess (as shown by the example of dashed green contours for 
$f_{\chi}=10^{-2}$). Hence, our model can readily explain the XENON1T excess without causing any tension with astrophysical observations.

% {\it Relic Density ---}
\section{Relic Density}
It is natural to ask how the light dark fermion becomes part of the DM and whether there exists parameter space that satisfies both the XENON1T 
experiment and the DM relic density~\cite{Planck2016AA}. From now on, we fix the dark fermion mass at the preferred value $m_{\chi}=3.17$~keV 
and the dark photon mass at $m_{\gamma'}=10^{-20}$~eV. If the dark fermions decouple from the thermal equilibrium in the early Universe when 
they are still relativistic, their current relic abundance is determined by 
$\Omega_{\chi}h^2=(4.67\times 10^{2}/g_{*}(T_f))(m_{\chi}/3.17~{\rm keV})$, 
where $g_{*}(T_f)$ is the number of relativistic degrees of freedom in the dark sector at the decoupling temperature $T_f$.  
A large value of $g_{*}\simeq 4\times 10^{3}$ is required to reproduce the DM relic density, which is fairly impossible in the current model.

In the freeze-out scenario, the dark fermions are initially in thermal equilibrium with other particles and decouple from the thermal bath when they become non-relativistic.  The solid curves in Fig.~\ref{fig:relic}(a) are for different relic densities of dark fermion produced via the freeze-out mechanism.  
Note that they have no dependence on the mixing parameter $\varepsilon$, as a result of the fact that the dark fermion annihilation to SM charged fermions is kinematically forbidden.  The dashed curves represent the best-fit value of ${\alpha}'$ in the $e_{X}$-$\varepsilon$ plane, with various choices of $f_{\chi}$ and the colored bands representing the 1$\sigma$ uncertainty.  
The intersection between the solid and dashed curves of the same color determines those parameters that satisfy the XENON1T excess while producing the correct fractional relic density of DM. 
We find that the stellar cooling from HB stars (grey region) has restricted the fraction $f_{\chi}$ to be in the range 
of $\gtrsim 1\%$ if the dark photon obtains its mass via the Higgs mechanism.  On the other hand, the stellar cooling constraints are 
insignificant for the SC dark photon with $m_{\gamma'}\lesssim 10^{-2}$~eV (pink region).  We should emphasize that the dark fermions are 
already cold when they freeze out at temperature $T\simeq m_{\chi}/10=0.32$~keV.  If the dark fermions from freeze-out constitute all of the 
DM relic, {\it i.e.} $f_{\chi}=1$, they can play the role of cold DM in the standard $\Lambda$CDM cosmology and do not alter the history of 
Universe structure formation beginning at $T\sim 10^{-2}$~eV.  We note in passing that as in the case of visible photon, the dark photon 
in the current scenario has a negligible relic density in the present Universe though it is the lightest stable particle in the dark sector.

If the dark fermions do not reach thermal equilibrium in the early Universe, their relic abundance can be obtained via 
the freeze-in mechanism~\cite{Hall2010JHEP}, as shown in Fig.~\ref{fig:relic}(b). 
This scenario requires $\varepsilon\gtrsim 10^{-7}$ and is thus excluded by the BBN constraint.

% {\it Summary ---}
\section{Summary}
We have proposed that the anomalous excess in the electron recoil events in XENON1T experiment can be owing to the effect of nearly massless dark 
photons produced from the annihilation of keV dark fermion in the Galaxy, and the dark fermion can constitute part of or entirely the dark matter 
relic density, depending on the parameters and scenario.  In particular, the required parameters for explaining the XENON1T excess and the DM 
relic density in the minimum model presented here are fully compatible with the constraints from stellar cooling.  
One of the attractive UV completed models for generating $\varepsilon\sim 10^{-11}-10^{-10}$ has been found in the realistic 
string theory in the large volume scenario~\cite{Goodsell2009JHEP}. Future multi-ton underground experiments shall be able to shed further 
light on the XENON1T anomaly and provide more opportunities to observe the dark sector.

{\it Acknowledgments---}
This work was supported in part by the Ministry of Science and Technology (MOST) of Taiwan under Grants No. 
MOST-108-2112-M-002-005-MY3 and No. MOST-108-2811-M-002-548.


\begin{thebibliography}{}

\bibitem{Planck2016AA} 
Planck Collaboration, Astron. Astrophys. {\bf 594}, A13 (2016).

\bibitem{Lee1977PRL} 
B. W. Lee and S. Weinberg, Phys. Rev. Lett. {\bf 39}, 165 (1977).

\bibitem{XENON1T2018} 
XENON Collaboration, Phys. Rev. Lett. {\bf 121}, 111302 (2018).

\bibitem{Ackermann2015PRL}
Fermi LAT Collaboration, Phys. Rev. Lett. {\bf 115}, 231301 (2015).

\bibitem{XENON1T2020} 
XENON Collaboration, Phys. Rev. D {\bf 102}, 072004 (2020).

\bibitem{Bloch2020} 
I. M. Bloch, A. Caputo, R. Essig, D. Redigolo, M. Sholapurkar, and T. Volansky, arXiv:2006.14521.

\bibitem{DFSZ} 
M. Dine, W. Fischler and M. Srednicki, Phys. Lett. B {\bf 104}, 199 (1981);
A. Zhitnitskii, Sov. J. Nucl. Phys. {\bf 31}, 260 (1980).

\bibitem{KSVZ} 
J. E. Kim, Phys. Rev. Lett. {\bf 43}, 103 (1979);
M. Shifman, A. Vainshtein and V. Zakharov, Nucl. Phys. B {\bf 166}, 493 (1980).


\bibitem{Alvarez2020} 
G. Alonso-Alvarez, F. Ertas, J. Jaeckel, F. Kahlhoefer, and L. J. Thormaehlen, arXiv:2006.11243;
H. An, M. Pospelov, J. Pradler, and A. Ritz, arXiv:2006.13929.

\bibitem{AcceleratedDM}
K. Agashe, Y. Cui, L. Necib, and J. Thaler, J. Cosmol. Astropart. Phys. {\bf 10}, 062 (2014);
D. Kim, J.-C. Park, and S. Shin, Phys. Rev. Lett. {\bf 119}, 161801 (2017);
G. F. Giudice, D. Kim, J.-C. Park, and S. Shin, Phys. Lett. B {\bf 780}, 543 (2018);
H. An, M. Pospelov, J. Pradler and A. Ritz, Phys. Rev. Lett. {\bf 120}, 141801 (2018);
T. Bringmann and M. Pospelov, Phys. Rev. Lett. {\bf 122}, 171801 (2019).

\bibitem{Alhazmi2020}
H. Alhazmi, D. Kim, K. Kong, G. Mohlabeng, J.-C. Park, and S. Shin, arXiv:2006.16252.

\bibitem{Fornal2020}
B. Fornal, P. Sandick, J. Shu, M. Su, and Y. Zhao, Phys. Rev. Lett. {\bf 125}, 161804 (2020).

\bibitem{Fabbrichesi2020} 
M. Fabbrichesi, E. Gabrielli, and G. Lanfranchi, arXiv:2005.01515.

\bibitem{NFW1997} 
J. F. Navarro, C. S. Frenk, and, S. D. M. White, Astrophys. J. {\bf 490}, 493 (1997).

\bibitem{Armengaud2013JCAP}
EDELWEISS Collaboration, J. Cosmol. Astropart. Phys. {\bf 11} (2013) 067.

\bibitem{PhotoelectricCS} 
W. M. J. Veigele, Atomic Data Table {\bf 5}, 51 (1973);
M. J. Berger, {\it et al.}, XCOM: Photon cross sections database, {\tt http://www.nist.gov/pml/data/xcom/index.cfm}.

\bibitem{An2015PLB}
H. An, M. Pospelov, J. Pradler, and A. Ritz, Phys. Lett. B {\bf 747} 331 (2015).

\bibitem{XENON2019Nature} 
XENON Collaboration, Nature (London) {\bf 568}, 532–535 (2019).

\bibitem{Vogel2014JCAP} 
H. Vogel and J. Redondo, J. Cosmol. Astropart. Phys. {\bf 02}, (2014) 029.

\bibitem{An2013PRL}
H. An, M. Pospelov, and J. Pradler, Phys. Rev. Lett. {\bf 111}, 041302 (2013).

\bibitem{McDermott2020PRD}
S. D. McDermott and S. J. Witte, Phys. Rev. D {\bf 101}, 063030 (2020).

\bibitem{Jaeckel2010PRD}
J. Jaeckel and S. Roy, Phys. Rev. D {\bf 82}, 125020 (2010).

\bibitem{Davidson1991PRD}
S. Davidson, B. Campbell, and D. Bailey, Phys. Rev. D {\bf 43}, 7 (1991);
S. Davidson, S. Hannestad and, G. Raffelt, J. High Energy Phys. {\bf 05} (2000) 003.

\bibitem{Signe2016AA} 
S. Riemer-S$\varnothing $rensen, Astron. Astrophys. {\bf 590}, A71 (2016).

% \bibitem{An2013PLB}
% H. An, M. Pospelov and J. Pradler, Phys. Lett. B {\bf 725}, 190 (2013).

\bibitem{Barducci2018CPC}
D. Barducci {\it et al.}, Comput. Phys. Commun. {\bf 222}, 327 (2018).

\bibitem{Hall2010JHEP} 
L. J. Hall, K. Jedamzik, J. March-Russell and S. M. West, J. High Energy Phys. {\bf 03}, 080 (2010).

\bibitem{Goodsell2009JHEP}
M. Goodsell, J. Jaeckel, J. Redondo and A. Ringwald, J. High Energy Phys. {\bf 11} (2009) 027.


\end{thebibliography}
\end{document}